\documentclass[twocolumn,aps,showpacs,prb,epsf,graphics,psfig]{revtex4}
\usepackage{graphicx}
\usepackage{graphicx}
\usepackage{bbold}
\usepackage{color}
\usepackage[normalem]{ulem}
\usepackage{amssymb}
\usepackage{tikz}
\usepackage[version=3]{mhchem} 
\usepackage{amsmath}

\begin{document}
\title{The effect of non-ionizing excitations on the diffusion of ion species and inter-track correlations in FLASH ultra-high dose rate radiotherapy}
\author{Ramin Abolfath$^{1,\dagger}$, Alexander Baikalov$^{2,3}$, Stefan Bartzsch$^{2,4}$, Niayesh Afshordi$^{5}$, Radhe Mohan$^{1}$}
\affiliation{
$^1$Department of Radiation Physics and Oncology, University of Texas MD Anderson Cancer Center, Houston, TX, 75031, USA \\
$^2$ Technical University of Munich, Department of Physics, Garching, Germany \\
$^3$ Helmholtz Zentrum M\"unchen GmbH, German Research Center for Environmental
Health, Institute of Radiation Medicine, Neuherberg, Germany \\
$^4$ Technical University of Munich, School of Medicine and Klinikum rechts der Isar,
Department of Radiation Oncology, Munich, Germany \\
$^5$ Department of Physics, University of Waterloo, Waterloo, Canada
}



\date{\today}
\begin{abstract}
{\bf Purpose}:
We present a microscopic mechanism that accounts for the outward burst of ``cold" ion species (IS) in a high-energy particle track due to coupling with ''hot" non-ion species (NIS). IS refers to radiolysis products of ionized molecules, whereas NIS refers to non-ionized excitations of molecules in a medium. The interaction is mediated by a quantized field of acoustic phonons, a channel that allows conversion of thermal energy of NIS to kinetic energy of IS, a flow of heat from the outer to the inner core of the track structure.

{\bf Methods}:
We perform step-by-step Monte Carlo (MC) simulations of ionizing radiation track structures in water to score the spatial coordinates and energy depositions that form IS and NIS at atto-second time scales.
We subsequently calculate the resulting temperature profiles of the tracks with MC track structure simulations and verify the results analytically using the Rutherford scattering formulation.
These temperature profiles are then used as boundary conditions in a series of multi-scale atomistic molecular dynamic (MD) simulations that describe the sudden expansion and enhanced diffusive broadening of tracks initiated by the non-equilibrium spectrum of high-energy IS.
We derive a stochastic coarse-grained Langevin equation of motion for IS from first-principle MD to describe the irreversible femto-second flow of thermal energy pumping from NIS to IS, mediated by quantized fields of acoustic phonons.
A pair-wise Lennard-Jones potential implemented in a classical MD is then employed to validate the results calculated from the Langevin equation.

{\bf Results}:
We demonstrate the coexistence of ``hot" NIS with ``cold" IS in the radiation track structures right after their generation.
NIS, concentrated within nano-scales volumes wrapping around IS, are the main source of intensive heat-waves and the outward burst of IS due to femto-second time scale IS-NIS coupling.
By comparing the transport of IS coupled to NIS with identical configurations of non-interacting IS in thermal equilibrium at room temperature, we demonstrate that the energy gain of IS due to the surrounding hot nanoscopic volumes of NIS significantly increases their effective diffusion constants.
Comparing the average track separation and the time scale calculated
for a deposited dose of 10 Gy and a dose rate of 40 Gy/s, typical values used in FLASH ultra high dose rate (UHDR) experiments, we find that the sudden expansion of tracks and ballistic transport proposed in this work strengthens the hypothesis of inter-track correlations recently introduced to interpret mitigation of the biological responses at the FLASH-UHDR~[\onlinecite{Abolfath2020:MP}].

{\bf Conclusions}:
The much higher diffusion constants predicted in the present model suggest higher inter-track chemical reaction rates at FLASH-UHDR, as well as lower intra-track reaction rates.
This study explains why research groups relying on the current Monte Carlo frameworks have reported negligible inter-track overlaps, simply because of underestimation of the diffusion constants.
We recommend incorporation of the IS-NIS coupling and heat exchange in all MC codes to enable these tool-kits to appropriately model reaction-diffusion rates at FLASH-UHDR.
\end{abstract}

\maketitle
\section{Introduction}
The unique normal tissue sparing of FLASH ultra high dose rates (UHDR), i.e., 40 Gy/s or higher, has recently garnered considerable attention
[\onlinecite{Favaudon2014:STM,Montay-Gruel2018:RO,Vozenin2018:CCR,Montay-Gruel2019:PNAS,Buonanno2019:RO,Vozenin2019:RO,Arash2020:MP,Spitz2019:RO,Koch2019:RO,Abolfath2020:MP,Seco2021:MP}].
The interpretation of the experimental data and the underlying microscopic mechanism is, however, under investigation and debated among researchers in the field of radiation therapy.

Passage of high energy particles and radiation in cells, tissues, or water-equivalent materials induces ionizations in addition to non-ionizing molecular excitations, in a hierarchical and complex process known as water radiolysis, with the following characteristics:
(1) ion species (IS, products of ionized water molecules) form a local plasma and react chemically with biological sensitive sites, e.g., DNA and chromosomes, and induce cell damage and eventually cell death (2) non-ionizing excitations (NIS) in the form of molecular vibrations, rotations and inter-orbital electronic excitations create highly localized hot volumes with nanoscopic lateral scales that wrap around IS.

Among all products of water radiolysis, the production of electrons, $e^-$, hydrogen atoms, $\ce{^{.}H}$, and hydroxyl free radicals, $\ce{^{.}OH}$, via the reactions $\gamma + \ce{H_2O \rightarrow ^{.}OH + ^{.}H}$ and $\gamma + \ce{H_2O \rightarrow H_2O^+ + e^-}$ are the most prominent.
In a subsequent process, $e^-$ and four water molecules form an aqueous electron complex, $\ce{e^-_{\rm aq}}$.
Hydroxyl free radicals, $\ce{^{.}OH}$, are neutrally charged molecules with an unpaired magnetic moment (spin-1/2), responsible for almost 80\% of DNA damage~[\onlinecite{Abolfath2009:JPC}].

The mechanism of radiation-induced oxygen depletion that scavenges $\ce{^{.}H}$ and $\ce{e^-_{\rm aq}}$ through the reactions $\ce{O_2 + ^{.}H \rightarrow ^{.}HO_2}$ and $\ce{O_2 + e^-_{\rm aq} \rightarrow ^{.}O_2^-}$ has been proposed to explain the observed FLASH-UHDR sparing effects.
At conventional dose rates (CDR), these reactions are identified to hinder neutralization of $\ce{^{.}OH}$-radicals
as the oxygen depletion processes compete with the reactions $\ce{^{.}OH + ^{.}H \rightarrow H_2O}$ and $\ce{^{.}OH + e^-_{\rm aq} \rightarrow OH^-}$ by lowering the population of $\ce{^{.}H}$ and $\ce{e^-_{\rm aq}}$.
The oxygen depletion hypothesis infers the underlying mechanism for the well-known oxygen effect in radiobiology:
the presence of oxygen in a tissue, e.g., at physoxic levels ($\approx$ 4-5\% oxygen in normal tissues), contributes to elevated radio-sensitivity.
In contrast, tissues at hypoxic levels (e.g., $<$ 0.3\% in parts of tumors) show lower radiosensitivity.
In addition, the secondary reactions $\ce{^{.}OH + ^{.}HO_2 \rightarrow O_2 + H_2O}$ and $\ce{^{.}OH + ^{.}O_2^- \rightarrow O_2 + OH^-}$ consume $\ce{^{.}OH}$-radicals; hence, while these reactions recycle the depleted oxygen to the environment, they hinder the reactivity of $\ce{^{.}OH}$-radicals.
The peak of oxygen depletion occurs on the micro-second time scale after the initial ionization.

At FLASH-UHDR, a self-recombination of $\ce{e^-_{\rm aq}}$ and $\ce{^{.}H}$ through the processes
$\ce{e^-_{\rm aq} + e^-_{\rm aq} + 2H_2O \rightarrow H_2 + 2OH^-}$ and $\ce{^{.}H + ^{.}H \rightarrow H_2}$
has been proposed as a competing mechanism to the aforementioned oxygen depletion mechanism~[\onlinecite{Seco2021:MP}].

Recently, a much faster process that directly blocks the reactivity of $\ce{^{.}OH}$-radicals and their tendency to cause DNA-damage has been proposed~[\onlinecite{Abolfath2020:MP}] as an alternative channel of self-interacting  $\ce{^{.}OH}$-radicals at FLASH-UHDR.
A series of molecular dynamics simulations have shown the combination of $\ce{^{.}OH}$-radicals in the form of stable compounds such as hydrogen peroxide ($\ce{^{.}OH + \ce{^{.}OH} \rightarrow  H_2O_2}$) or transient complexes of $\ce{^{.}OH}$ with short life times. These self-interaction processes are responsible for lowering the population of reactive $\ce{^{.}OH}$-radicals, and is dominant in the sub-nano-second time scale. As dose rate increases from CDR to FLASH-UHDR, such self-interaction processes become more prevalent due to increased inter-track overlap.

The track-origin of self-interacting particles, and whether pairs of IS from the list $\ce{e^-_{\rm aq}}$, $\ce{^{.}H}$, or $\ce{^{.}OH}$, $\cdots$ are from a single track or a pair of tracks is key to understanding the differentiate biological responses of CDR and FLASH-UHDR.

Here we present a physical picture of IS-NIS coupling via a quantized field of acoustic phonons that justifies the multi-track origin of particle pairing at FLASH-UHDR.
The cold IS of a track gain kinetic energy from hot NIS and travel much farther from the core of the track than if they were to diffuse away at room temperature.
This outward burst of IS has a profound effect on the kinetics of intra- and inter-track chemical reactions, particularly on the reaction rates among reactive oxygen species (ROS) and formation of non-reactive oxygen species (NROS) at FLASH-UHDR~[\onlinecite{Abolfath2020:MP}].

In the current MC models developed for the nanoscopic studies of biological responses, including
Geant4~[\onlinecite{Agostinelli2003:NIMA}], Geant4-DNA~[\onlinecite{Incerti2010:IJMSSC}], TOPAS-nBio~[\onlinecite{Schuemann2019:RR,Faddegon2020:PM}], PARTRAC~[\onlinecite{Friedland2011:MR}], RITRACKS~[\onlinecite{Plante2011:RPD}] and gMicroMC~[\onlinecite{Lai2021:PMB}], the diffusion constants of chemical species (IS) are modeled phenomenologically, taken from experimental data with no consideration of the interactions between IS and NIS. The diffusive broadening of track-structures in these MC computational platforms is determined by the room-temperature diffusion constants of the individual species, which are proportional to their mean square displacements (Einstein relation).
In this work, we demonstrate that the lack of coupling between NIS and IS underestimates (overestimates) the inter-track (intra-track) correlations and hence wrongly predicts overall radiochemical yields.

We emphasize that fitting the current experimental data, with the hypotheses developed based on inter-track coupling of IS can turn to a paradigm shift to interpret and understanding
the FLASH-UHDR effects.
Thus, a shift in intra-track broadening owing to the irreversible flow of energy from hot NIS to cold IS may enhance inter-track coupling significantly, and hence the interpretation and analysis of the experimental data.
We therefore conclude that the current MC models must be corrected for the initial burst of the tracks to become suitable computational tools for the simulation of radiobiological responses, in particular the FLASH-UHDR effects.

The current study aims to address these effects by calculating
(a) the temperature profile of NIS
(b) the time scale of energy exchange between NIS and IS
(c) the length scale of non-equilibrium expansion of tracks until their transition to diffusive expansion under Brownian motion at room temperature.

We have organized the technical sections of our study in the following order:
a brief description of step by step energy deposition using MC is given in Sec. \ref{SecG4DNA}.
Calculation of temperature profiles of NIS for a range of energies is given in Sec. \ref{SecTprofile}.
Construction of an MD system of IS-NIS interacting via quantized fields of acoustic phonons is given in Secs. \ref{SecHam}-\ref{SecEqMotion}.
Construction of a stochastic coarse-grained Langevin equation, its boundary conditions, and calculation of the numerical solutions that provide the heat transfer parameters from NIS to IS is given in Sec. \ref{SecLangevin}.
Validation of the results obtained from this Langevin equation, based on atomistic simulation of a collection of atoms interacting by a Lennard-Jones potential, is given in Sec. \ref{SecLJ}.
Finally we wrap up our presentation with the results and conclusion in Secs. \ref{SecRes} and \ref{SecConclu} respectively.


\begin{figure}
\begin{center}
\includegraphics[width=0.8\linewidth]{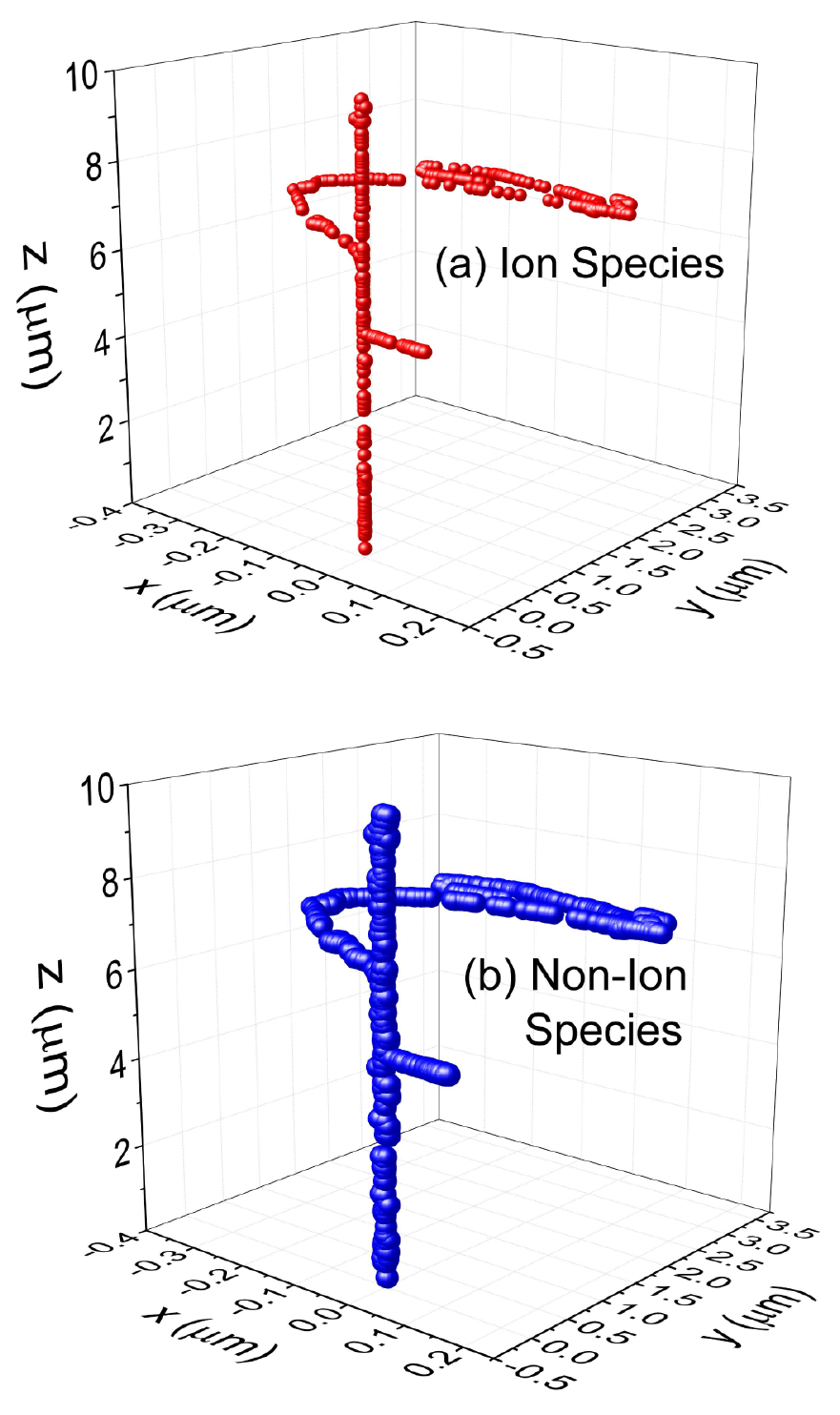}\\ 
\noindent
\caption{
Initial distribution of ionizing (top)
and non-ionizing (bottom) excitations of a single track of 80 MeV proton passing through water-equivalent medium within atto-second time frame calculated by step-by-step MC transport model of ionizing-radiation using Geant4-DNA toolkits.
}
\label{Fig1}
\end{center}\vspace{-0.5cm}
\end{figure}

\section{Method}
Our method of calculating the diffusive range of IS consists of a combination of numerical and analytical multi-scale computational approaches.
Calculation of step-by-step energy deposition and positions of IS and NIS in water were performed using a MC approach with Geant4-DNA.
The temperature profiles of NIS for a range of proton energies were obtained after ensemble averaging of these energy depositions in Geant4-DNA, and were validated analytically using a Rutherford scattering formulation.
Interactions between IS and NIS via quantized fields of acoustic phonons were constructed using a molecular dynamics model followed by a stochastic coarse-grained Langevin equation with a boundary condition obtained from the temperature profiles of NIS. The numerical solutions of the range of IS using the Langevin equation were performed, provided that the irreversible heat transfer from NIS to IS initially expands tracks to a point in space and time at which they continuously undergo a cooling-down transition under non-equilibrium conditions until they eventually relax to a thermal equilibrium condition at room temperature where the subsequent expansion follows a Gaussian distribution function.

\section{Geant4-DNA MC simulation}
\label{SecG4DNA}
For the sake of demonstration, in this work we focus on passage of a scanning beam of protons within a range of energies relevant to clinical applications of proton therapy, 80-250 MeV.

Figures \ref{Fig1}(a) and (b) illustrate the typical geometrical distribution of IS and NIS generated by a single track of an 80 MeV proton.
Each dot represents a point of ionization or excitation with $(x,y,z)$ coordinates calculated in water-equivalent medium by the Geant4-DNA MC toolkit~[\onlinecite{Incerti2010:IJMSSC}].
The non-ionizing molecular states scored in this MC simulation include vibrational, rotational and inter-orbital electronic excitations of water molecules.

Although the track structure of these two species overlap in space at the time of their creation, their thermodynamic characteristics are dramatically different.
Their local temperature is the most significant disparity among these two species, as
IS consist of cold (room temperature) molecules while NIS are hot with an average temperature up to orders of magnitude greater than room temperature.
There is also a disparity in their spatial extent: NIS are generated at longer length scales and hence wrap around IS, simply because the energy threshold of NIS creation is an order of magnitude lower than that of IS.

Due to their close proximity, it is expected that the flow of thermal energy from NIS to IS takes place immediately after their creation.
IS with a resulting gain in their kinetic energies may travel much longer distances compared to otherwise identical IS that move only room temperature Brownian motion.
Similar phenomena in the context of lightening have been studied for a long time (see for example Ref. [\onlinecite{Dong2016:AR}]).

Here we present some interesting statistics from an ensemble average of the simulation shown in Fig.~\ref{Fig1}.
We have scored 204 ionizations in a nucleus of a cell with diameter $5\mu m$, and thus an average of approximately 40 ions per $ 1\mu m$.
If, on average, each ionization accounts for $13 eV$ of energy deposition, then approximately $2700 eV$ are deposited in a cell nucleus.
Conversion to a linear-energy transfer can be done easily: $2700 eV/5\mu m = (2.7/5) keV/\mu m = 0.54 keV/\mu m$.
This is in contrast to NIS, as we have scored $40,000$ excitation points with a total energy of $1200 eV$; almost 40\% of the total energy loss to IS.

The number of IS and NIS events increases with decreasing proton energy.
For example, a 100 keV proton yields approximately 4400 ions in a volume with a diameter of $2\mu m$, as the range of a proton with an initial energy of 100 keV is limited to $2\mu m$.
This is equivalent to 2200 ionizations per $\mu m$.
The total deposited energy is 55,000 eV, equivalent to a linear energy transfer of 55 keV/(2 $\mu m$) = 27.5 keV/$\mu m$.
Therefore, from a simple energy balance, one can infer the rest of the proton energy (45 keV) must be spent to generate excitations in the form of heat, i.e., NIS.
A 100keV proton yields approximately 900,000 excitation points (NIS).

As protons pass through matter, the energy loss, in the form of both ionizations and thermal excitations, per unit length increases.
Thus, an increase in local temperature is expected as the energy of the proton decreases.
As shown in Fig. \ref{fig:Tprofile}, the highest local temperature is predicted corresponding to a proton with an optimal kinetic energy.

\section{Temperature profiles of ionizing tracks}
\label{SecTprofile}
In this section, we calculate the radial temperature profile of the track of a fast ion of velocity $V_I$, charge $Z_I e$ and mass $m_I$ via impulse approximation.
From Rutherford scattering, we know that any particle $a$ of charge $Z_a e$ and mass $m_a$ on the path of a fast ion has the scattering angle $\Theta$ in the rest frame of the fast ion, where
\begin{equation}
    \tan(\Theta/2) = \frac{Z_I Z_a e^2(m_a+m_I)}{m_a m_I V^2_I b},
\end{equation}
$b$ is the impact parameter, and we ignore the original velocity of the particle $a$. In the lab frame, the kinetic energy of particle $a$ after scattering is:
\begin{equation}
    E_a = m_a V^2_I (1- \cos\Theta) = \frac{2m_a V^2_I}{1+ \left[\frac{m_a m_I V^2_I b}{Z_I Z_a e^2(m_a+m_I)}\right]^2}.
\end{equation}
While this is a classical result, we can include quantum effects by noting that the angular momentum in the center of mass frame is quantized in units of $\hbar$, and thus
\begin{equation}
    \frac{m_a m_I b V_I}{(m_a+m_I)} \gtrsim \hbar.
\end{equation}
We can impose this minimum by adding a term in the denominator, which effectively limits how small $b$ can get:
\begin{equation}
   E_a \simeq \frac{2m_a V^2_I}{1+ \left[\frac{m_a m_I V^2_I b}{Z_I Z_a e^2(m_a+m_I)}\right]^2 + \left( V_I \hbar \over Z_I Z_a e^2\right)^2  }.
\end{equation}

The next correction comes from the limitation of the impulse approximation. The momentum kick received by a particle in a harmonic oscillator with frequency $\omega_a$ is given by
\begin{equation}
    p_a = Z_I Z_a e^2 b\int_{-\infty}^\infty dt \frac{\exp(i \omega_a t)}{[(V_I t)^2+b^2]^{3/2}}.
\end{equation}
From this, we can see that the impulse approximation that assumes $\omega_a \ll V_I/b$ will receive a correction:
\begin{equation}
\frac{p_a}{p_a(\omega_a = 0)} = \frac{b\omega_a}{V_I}K_1\left(\frac{b\omega_a}{V_I} \right),
\end{equation}
where $K_1(x)$ is the modified Bessel function of the second kind, and asymptotically falls to $\sqrt{\pi x/2} \exp(-x)$ as $x\to \infty$. Here, $\omega_a$ represents atomic frequencies. For external forces that vary more slowly than $\omega_a$, the atoms react adiabatically and do not get excited. While lower frequency molecular modes can still be excited, they only couple to higher multipoles, which is further suppressed. Therefore, our final formula (given that $E_a \propto  p^2_a$) is:
\begin{equation}
   E_a(b) \simeq  \frac{2m_a b^2 \omega^2_a K^2_1(b \omega_a /V_I)}{1+ \left[\frac{m_a m_I V^2_I b}{Z_I Z_a e^2(m_a+m_I)}\right]^2 + \left( V_I \hbar \over Z_I Z_a e^2\right)^2  }.
\end{equation}

\begin{figure}
\centering
\includegraphics[width=0.4\textwidth]{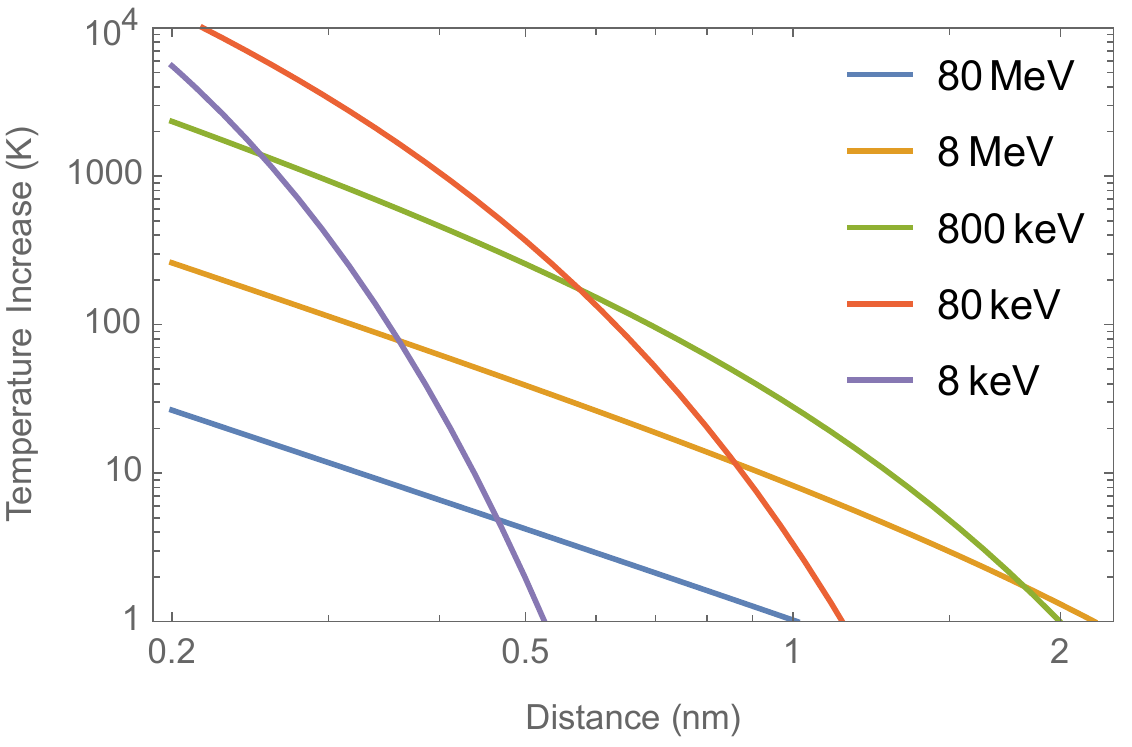}
\caption{
Temperature increase profiles of water molecules due to passage of a proton with energies $E_I = \frac{1}{2} m_p V^2_I$ = 8 keV-80 MeV.
\label{fig:Tprofile}
}
\end{figure}

Finally, the increase in temperature is given by:
\begin{equation}
    \rho c_v \Delta T(b) = \sum_a n_a E_a(b),
\end{equation}
where $n_a$ is the number density of particle species $a$, $\rho$ is the medium density, and $c_v$ is the specific heat capacity at constant volume. Figure \ref{fig:Tprofile} shows the resulting temperature increase profiles for water molecules as a result of the passage of a proton with energies ranging from 80 MeV to 8 keV.

\section{Heat transfer from NIS to IS: a molecular dynamics formulation}
\subsection{Hamiltonian}
\label{SecHam}
In this section we compose a model to calculate the rate of heat exchange from hot NIS to cold IS.
The molecular dynamics model developed in this section constitutes equations previously employed for transition path sampling study of classical rate-promoting vibrations in enzymatic reactions [\onlinecite{Antoniou2004:JCP}].
Throughout this formulation, we calculate macroscopic transport coefficients such as the diffusion constant.
In addition, we calculate the energy dissipation rate, $\xi$, and the fluctuating field, $\delta F$, a pair of parameters that govern the dissipation-fluctuation theorem in a system of many particles~[\onlinecite{Girvin:Book}].

We represent an IS with a classical particle (the system) interacting with a number of particles that form a thermal bath (NIS, the environment) as shown in Fig.~\ref{Fig3_IS_NIS_coupling}.

\begin{figure}
\begin{center}
\includegraphics[width=0.8\linewidth]{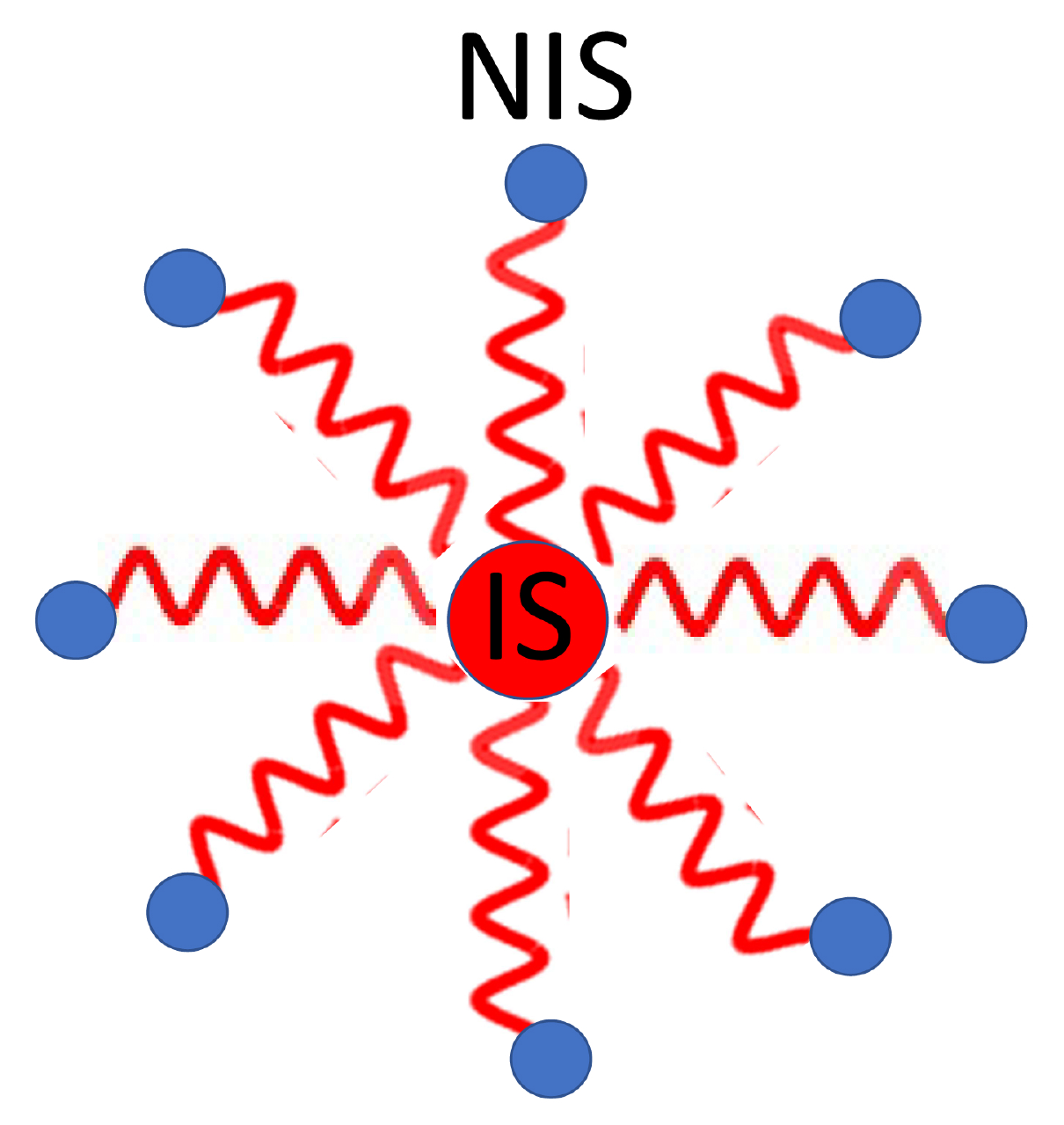}\\ \vspace{-0.5cm}
\noindent
\caption{
Sketch of IS-NIS coupling. The wiggly lines represents the coupling through phonon fields.
}
\label{Fig3_IS_NIS_coupling}
\end{center}\vspace{-0.5cm}
\end{figure}

These particles can be represented by a Hamiltonian system of classical harmonic oscillators that vibrate around their minimum energies because of the thermal energy they contain.
The whole thermal energy is conserved and can be transferred to the particle via exchange coupling.

Thus, the total Hamiltonian can be partitioned into a Hamiltonian of the system (IS), $H_S$, the thermal bath (NIS), $H_B$, and the exchange interaction between the system and the bath, $V$
\begin{eqnarray}
H = H_S + H_B + V.
\label{eq11}
\end{eqnarray}
Here
\begin{eqnarray}
H_S = \frac{\vec{p}^2}{2m} + \frac{1}{2} m \omega_s \vec{r}^2 + U(\vec{r}),
\label{eq12}
\end{eqnarray}
where $\vec{p}$, $\vec{r}$ and $m$ represent the momentum, position and rest mass of IS, respectively.
$\omega_s$ is the frequency of a simple harmonic potential that the particle experiences.
$U(\vec{r})$ represents an external potential energy the particle experiences.
For a free IS, in the absence of any external electric and magnetic fields, $U=0$,
and
\begin{eqnarray}
H_B = \sum_{j=1}^{N_B} \left[\frac{\vec{p}^2_j}{2} + \frac{1}{2}\omega_j \vec{q}^2_j\right].
\label{eq13}
\end{eqnarray}
For the sake of simplicity in calculation we assume that an ensemble of particles in the thermal bath carries a unit rest mass.
In Eq.(\ref{eq13}), $\vec{p}_j$ and $\vec{q}_j$ represent the momentum and position of the $j$-th particle in the thermal bath.
$\omega_j$ is the associated simple harmonic oscillator frequency.
The linear coupling between the system and the bath can be written as
\begin{eqnarray}
V = - \left(\sum_{j=1}^{N_B} \frac{\gamma_j}{\omega_j} \vec{q}_j\right) \cdot\vec{r},
\label{eq14}
\end{eqnarray}
where $\gamma_j$ is the strength of coupling between the $j$-th NIS particle in the bath and IS particle in the system.
Without loss of generality, one can consider the following specific relation
\begin{eqnarray}
\omega_s = \sum_{j=1}^{N_B} \frac{\gamma^2_j}{\omega^3_j}
\label{eq15}
\end{eqnarray}
that allows for a compact form of the total Hamiltonian:
\begin{eqnarray}
H &=& H_S + H_B + V \nonumber \\
&=&
\frac{\vec{p}^2}{2m} + U(\vec{r})
+ \sum_{j=1}^{N_B} \left[\frac{\vec{p}^2_j}{2} + \frac{1}{2}\omega_j
\left(\vec{q}_j - \frac{\gamma_j}{\omega^2_j} \vec{r}\right)^2\right].\nonumber \\
\label{eq16}
\end{eqnarray}

\subsection{Equations of motion}
\label{SecEqMotion}
The classical equations of motion can be obtained from variation of the Hamiltonian with respect to the coordinates and momenta of the IS particles in the system and NIS particles in the thermal bath.
The Hamilton equations for the particle in the system are given by
\begin{eqnarray}
\dot{\vec{r}}(t) = \frac{\delta H}{\delta \vec{p}(t)} = \frac{\vec{p}(t)}{m},
\label{eq17}
\end{eqnarray}
and
\begin{eqnarray}
\dot{\vec{p}}(t) &=& - \frac{\delta H}{\delta \vec{r}(t)}
= -\vec{\nabla}U(\vec{r}) + \sum_{j=1}^{N_B} \gamma_j \left(\vec{q}_j(t) - \frac{\gamma_j}{\omega^2_j}\vec{r}(t)\right).
\nonumber \\
\label{eq18}
\end{eqnarray}
The equations of motion for the particles in the thermal bath are given by
\begin{eqnarray}
\dot{\vec{q}}_j(t) = \frac{\delta H}{\delta \vec{p}_j(t)} = \vec{p}_j(t),
\label{eq19}
\end{eqnarray}
and
\begin{eqnarray}
\dot{\vec{p}}_j(t) = - \frac{\delta H}{\delta \vec{q}_j(t)}
= - \omega^2_j \vec{q}_j(t) + \gamma_j \vec{r}(t).
\label{eq20}
\end{eqnarray}
The last term in Eq.(\ref{eq20}) resembles an external force.
Hence Eqs. (\ref{eq19}) and (\ref{eq20}) describe a system of harmonic oscillators in an external force, acting on the particles in the bath.
The solutions of this equations can be calculated using the Green's function method (see page 401, Ref.[\onlinecite{ByronFuller:Book}])
\begin{eqnarray}
\vec{q}_j(t) &=& \vec{q}_j(0) \cos\omega_j t + \vec{p}_j(0)\frac{\sin\omega_j t}{\omega_j} \nonumber \\
&+& \gamma_j \int_{0}^{t} dt' \frac{\sin\omega_j (t-t')}{\omega_j} \vec{r}(t'),
\label{eq21}
\end{eqnarray}
where $G(t-t') = \sin\omega_j (t-t')/\omega_j$ is the Green's function of a forced simple harmonic oscillator.
Performing integration by part converts Eq.(\ref{eq21}) to
\begin{eqnarray}
\vec{q}_j(t) &-& \frac{\gamma_j}{\omega^2_j} \vec{r}(t) =
\left[\vec{q}_j(0) - \frac{\gamma_j}{\omega^2_j} \vec{r}(0)\right] \cos\omega_j t \nonumber \\
&+& \vec{p}_j(0) \frac{\sin\omega_j t}{\omega_j}
- \gamma_j \int_{0}^{t} dt' \frac{\cos\omega_j (t-t')}{\omega^2_j} \frac{\vec{p}(t')}{m}. \nonumber \\
\label{eq22}
\end{eqnarray}
Considering $U=0$ and inserting Eq.(\ref{eq22}) in Eq.(\ref{eq18}),
we find a non-local Langevin equation of motion for the ion-species
\begin{eqnarray}
\dot{\vec{p}}(t) = - \frac{1}{m}\int_{0}^{t} dt' K(t-t') \vec{p}(t') + \delta \vec{F}(t),
\label{eq23}
\end{eqnarray}
where the kernel of the integral, $K(t-t')$, is a ``memory-function"
\begin{eqnarray}
K(t-t') &=& \sum_{j=1}^{N_B} \frac{\gamma_j}{\omega^2_j} \cos\omega_j (t-t') \nonumber \\
&=& \int_{0}^{\infty} d\omega g(\omega) \frac{\gamma^2(\omega)}{\omega^2} \cos\omega (t-t').
\label{eq24}
\end{eqnarray}
Here, $g(\omega)$ is the density of states (DOS) of the thermal bath degrees of freedom in the frequency domain.
The transformation $\sum_{j=1}^{N_B} \rightarrow \int_{0}^{\infty} d\omega g(\omega)$
has been performed, where $g(\omega) = \sum_{j=1}^{N_B} \delta(\omega - \omega_j)$.

A specific form of the DOS can be found, $g(\omega) = g_0 \omega^2$, where $g_0 = 1/(2\pi^2 c^3)$.
Here $\omega = c k$ is the dispersion relation of the quantized field of acoustic phonons, which are the normal modes of NIS.
$c$ is the speed of sound in medium, i.e., thermal bath or water in our case.
Assuming a non-dispersive coupling constant, $\gamma(\omega) = \gamma_0$, yields a delta-function type of memory function that is local in time with a time-evolution process in Eq.(\ref{eq23}), equivalent to a Markov-chain
\begin{eqnarray}
K(t-t') = \xi \delta(t - t').
\label{eq25}
\end{eqnarray}
Here $\xi$ is the strength of the damping or friction force,
\begin{eqnarray}
\xi = \gamma_0 g_0 = \frac{\gamma_0}{2\pi^2 c^3},
\label{eq25_1}
\end{eqnarray}
hence
\begin{eqnarray}
\omega_s &=& \sum_{j=1}^{N_B} \frac{\gamma^2_j}{\omega^3_j}
= \int_{0}^{\infty} d\omega g(\omega)\frac{\gamma^2(\omega)}{\omega^3} \nonumber \\
&=&  \int_{0}^{\infty} d\omega (g_0 \omega^2) \frac{\gamma^2_0}{\omega^3}
= g_0 \gamma^2_0 \int_{0}^{\infty} d\omega \frac{1}{\omega} \nonumber \\
&=& g_0 \gamma^2_0 \left. \ln(\omega) \right|_{0}^\infty
\label{eq15_1}
\end{eqnarray}
Note that Eq.(\ref{eq23}) reduces to the local-in-time Langevin equation.
The random-force (noise), $\delta F(t)$, is given by the random configuration of particles in the bath at the initial time, $t=0$
\begin{eqnarray}
\delta \vec{F}(t) &=& \sum_{j=1}^{N_B} \gamma_j \vec{p}_j(0)\frac{\sin\omega_j t}{\omega_j} \nonumber \\
&+& \sum_{j=1}^{N_B} \left[\vec{q}_j(0) - \frac{\gamma_j}{\omega^2_j}\vec{r}(0)\right] \cos\omega_j t.
\label{eq26}
\end{eqnarray}
Here $\vec{p}_j(0)$, $\vec{q}_j(0)$, and $\vec{r}(0)$ are the initial momenta and positions of the particles.
If the bath has a large number of independent particles, then there would be no correlations among the initial
coordinates and momenta; thus $\delta \vec{F}$ is indeed a random variable.
For a bath at a given temperature $T$, a canonical ensemble average over the initial condition of the Hamiltonian $H$ can be performed through a Boltzmann partition function,
\begin{eqnarray}
Z = e^{-H/k_B T}.
\label{eq27}
\end{eqnarray}
Because $H = H(\vec{r}(0), \vec{p}(0), \{\vec{q}_{j}(0)\}, \{\vec{p}_{j}(0)\})$ is a quadratic function of position and momentum of particles, $Z$ is equivalent to a Gaussian distribution function over the random initial coordinates in phase space of both IS particles in the system and NIS particles in the thermal bath.
Moreover, because of ergodicity of the system and the thermal bath, time averaging can be replaced by equilibrium ensemble averaging such that
the ensemble average of the first moments of the particles' positions and momenta vanishes
\begin{eqnarray}
\langle \vec{q}_j(0) - \frac{\gamma_j}{\omega^2_j}\vec{r}(0)\rangle = 0,
\label{eq28}
\end{eqnarray}
and
\begin{eqnarray}
\langle \vec{p}_j(0) \rangle = 0.
\label{eq29}
\end{eqnarray}
The second moments are
\begin{eqnarray}
\left\langle \left[\vec{q}_j(0) - \frac{\gamma_j}{\omega^2_j}\vec{r}(0)\right]^2\right\rangle = \frac{k_B T}{\omega^2_j},
\label{eq30}
\end{eqnarray}
and
\begin{eqnarray}
\langle \vec{p}~^2_j(0) \rangle = k_B T.
\label{eq31}
\end{eqnarray}
From these equations one can show explicitly
\begin{eqnarray}
\langle\delta F_k(t) \delta F_{k'}(t')\rangle = k_B T K(t - t') \delta_{kk'},
\label{eq31}
\end{eqnarray}
where $k,k' = (x,y,z)$ are Cartesian components of the random force $\delta \vec{F}$.

\begin{figure}
\begin{center}
\includegraphics[width=0.8\linewidth]{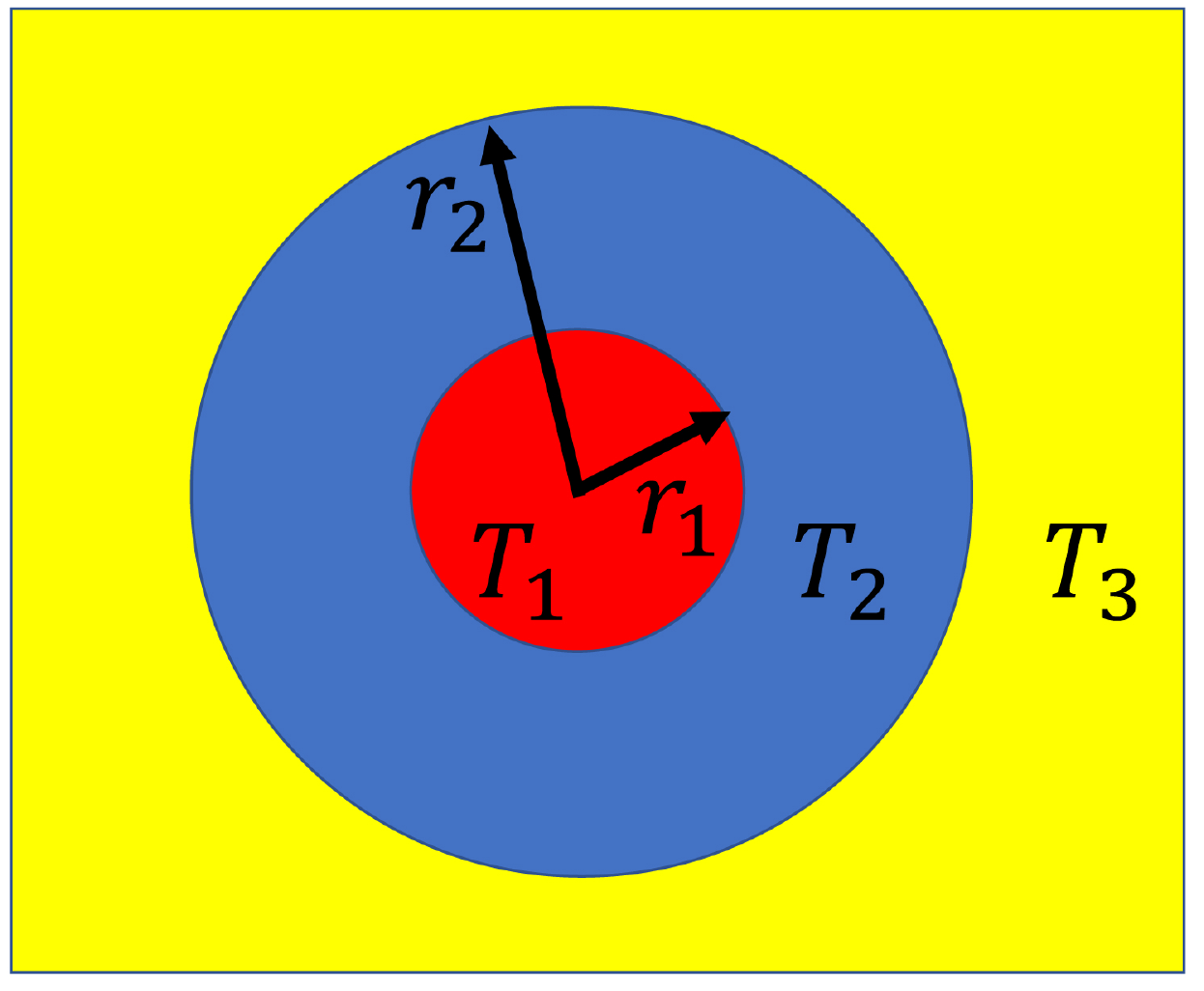}\\ 
\noindent
\caption{
Sketch of boundary conditions used to calculate the dynamic solutions of the Langevin equations.
The inner and outer cores of a track corresponds to IS and NIS, represented by red and blue circles with radii $r_1$ and $r_2$ and local temperatures $T_1$ and $T_2$, respectively. It should be noted that NIS with temperature $T_2$ are also located within the red circle with radius $r_1$.
The outside environment corresponding to $r > r_2$ and temperature $T_3$ is represented by the color yellow.
}
\label{Fig3_IS_NIS_coupling}
\end{center}\vspace{-0.5cm}
\end{figure}

\subsection{Langevin dynamics}
\label{SecLangevin}
To calculate the trajectory of an IS, e.g., H or OH, as discussed in Sec. (\ref{SecHam}), we engage a stochastic equation of motion that allows energy exchange between cold IS and hot NIS directly after their creation, using a boundary condition as illustrated in Fig. \ref{Fig3_IS_NIS_coupling}.
In particular, the solutions of a system of the Langevin equation, Eq.(\ref{eq23}), with a kernel given by Eq.(\ref{eq25}), corresponding to interactions mediated by a quantized field of harmonic oscillations, acoustic phonons, may serve the purpose of this study:
\begin{eqnarray}
m\frac{d\vec{v}(t)}{dt} = -\xi \vec{v}(t) + \delta\vec{F}(t).
\label{eq1}
\end{eqnarray}
Here $m$, $\vec{v}$, $\xi$ and $\delta\vec{F}$ are the IS rest mass, IS velocity, and dissipative force constant (friction) and a random force (noise) acting on IS, respectively.
$t$ is the time right after passage of a radiation beam that creates IS and NIS simultaneously.
The origin of the friction force and noise is the coupling with the ``thermal bath", i.e., molecules from non-ionizing excitations: NIS and the background water molecules.

The analytical solution of Eq.(\ref{eq1}) is given by
\begin{eqnarray}
\vec{v}(t) = \vec{v}_0 e^{-\xi t/m} + \frac{1}{m} \int_{0}^{t} dt' e^{-\xi (t - t')/m} \delta\vec{F}(t').
\label{eq4}
\end{eqnarray}
$\vec{v}_0$ is the initial velocity of the particle at $t=0$.

In a Markov random process (white noise) where
\begin{eqnarray}
\langle\delta\vec{F}(t)\rangle = 0,
\label{eq2}
\end{eqnarray}
there is no correlation in time, i.e., $\langle\delta F_k(t) \delta F_{k'}(t')\rangle = \langle\delta F_k(t)\rangle  \langle\delta F_{k'}(t')\rangle = 0$ if $t \neq t'$.
Here $k$ refers to vector components of the random force, $\delta \vec{F}$, in Cartesian space, $k=(x,y,z)$.
For $t = t'$, and $k=k'$ we have $\langle\delta F_k(t) \delta F_{k'}(t')\rangle = \langle\delta F^2_k(t)\rangle = 2B$ hence
\begin{eqnarray}
\langle\delta F_k(t) \delta F_{k'}(t')\rangle = 2B \delta(t - t') \delta_{k,k'}.
\label{eq3}
\end{eqnarray}
$B$ is a measure of the fluctuating force strength and $\delta(t - t')$ implies no temporal correlations.
Note that $\langle \cdots \rangle$ is the ensemble average over the initial state of the system and thermal bath (environment).
Eqs.(\ref{eq2}-\ref{eq3}) suggest a normalized Gaussian distribution function for the noise
\begin{eqnarray}
P(\delta F_k(t)) = \frac{1}{2\sqrt{\pi B}} e^{- \frac{\delta F^2_k(t)}{4B}}.
\label{eq3_1}
\end{eqnarray}

To connect the noise strength ($B$) to the thermal bath temperature, $T$, we calculate $\langle \vec{v}^2(t) \rangle$
\begin{eqnarray}
\langle \vec{v}^2(t) \rangle &=& \vec{v}^2_0 e^{-2\xi t/m}
+  \frac{2}{m} \vec{v}_0  \cdot e^{-2\xi t/m} \int_{0}^{t} dt' e^{-\xi t'/m} \langle\delta \vec{F}(t')\rangle \nonumber\\
&+& \frac{e^{-2\xi t/m}}{m^2} \int_{0}^{t} dt' \int_{0}^{t} dt'' e^{\xi \frac{t' + t"}{m}} \langle\delta\vec{F}(t')\cdot\delta\vec{F}(t'')\rangle \nonumber\\
&=& \vec{v}^2_0 e^{-2\xi t/m} + 3\frac{B}{\xi m} \left(1 - e^{-2t\xi/m}\right).
\label{eq7}
\end{eqnarray}
We insert Eqs.(\ref{eq2}-\ref{eq3}) to arrive at the last equality.
Considering the asymptotic limit of IS kinetic energy at long times \
leads to an identity known as the ``fluctuation-dissipation theorem"
\begin{eqnarray}
2B = k_B T \xi,
\label{eq8}
\end{eqnarray}
that relates the fluctuation strength, $B$, with the dissipation strength $\xi$, both originating from the interaction of the IS with the environment.
Here $T$ refers to room temperature.
This expression gives a balance between dissipation (friction), which tends to drive the system (IS) to a completely ``dead" state, and ``noise", which tends to keep the system alive.
This balance pushes the system to reach a state of thermal equilibrium at large enough times.

The integration of random forces in the second term of Eq. (\ref{eq4}) was performed numerically by splitting the time axis into a finite-size time-series within $0 \leq t' \leq t$.
In every time-step, a random force was drawn from the Gaussian distribution function, Eq.(\ref{eq3_1}).
To incorporate the thermal fluctuations in numerical solutions of Eq. (\ref{eq4}),
the temperature profiles, as given in Fig. \ref{Fig3_IS_NIS_coupling}, were
enforced as an input in the boundary conditions of Eq. (\ref{eq4}), i.e.,
a series of constraints in the broadening of the Gaussian random forces in Eq. (\ref{eq8}).
To this end, we assigned $(B_1, B_2, B_3) = k_B (T_1, T_2, T_3) \xi$ in the following domains $r \leq r_1$, $r_1 < r \leq r_2$ and $r > r_2$ in Eq.(\ref{eq3_1}), where $r$ is the lateral position of IS in cylindrical coordinates.

To calculate the diffusion constant, we use the Kubo formula and the velocity-velocity correlation function
\begin{eqnarray}
D &=& \frac{1}{3} \int_{0}^{\infty} dt \langle \vec{v}(t)\cdot \vec{v}(0) \rangle =
\frac{k_B T}{\xi}.
\label{eq9axd}
\end{eqnarray}
To arrive at the last identity in Eq.(\ref{eq9axd}), we apply Eq.(\ref{eq4}) and then Eq.(\ref{eq2}) in the first part of Eq.(\ref{eq9axd}).
From this equation one can calculate a numerical value for $\xi$.
Considering empirical values of the mean diffusion constant of IS in water, $D_{\rm eq} = 4.3 \times 10^{-9} \frac{m^2}{s}$, we find
\begin{eqnarray}
\xi &=& \frac{k_B T}{D_{\rm eq}} = \frac{1.38\times 10^{-23} m^2\cdot kg\cdot s^{-2}\cdot K^{-1} \times 300 K}{4.3\times 10^{-9} m^2 s^{-1}} \nonumber \\
&\approx& 100 \times 10^{-14} kg\cdot s^{-1} = 10^{-12} kg\cdot s^{-1}.
\label{eq51n}
\end{eqnarray}
where $k_B = 1.38064852 \times 10^{-23} m^2\cdot kg\cdot s^{-2}\cdot K^{-1}$ is the Boltzmann constant. Here $T=300 K$ is the room temperature.
It would be useful to calculate a time scale, $t_1 = \frac{m}{\xi}$, beyond which the system relaxes to a steady state.
By converting Eq.(\ref{eq51n}) to
\begin{eqnarray}
\frac{\xi}{m_H} &\approx& \frac{10^{-12} kg\cdot s^{-1}}{1.67 \times 10^{-27} kg} \nonumber \\
&\approx& 0.6 \times 10^{15} s^{-1} = 0.6 \times 10^{9} \mu s^{-1}
= 0.6fs^{-1},\nonumber \\
\label{eq52n}
\end{eqnarray}
and considering a Hydrogen atom as a typical IS where $m_H = 1.6735575 \times 10^{-27} kg$, we find $t_{1H} \approx 1 fs$.
Considering heavier species such as OH, one can find longer relaxation times for the KE transfer, e.g., $t_{1OH} \approx 17 fs$.
Based on the IS rest mass, the typical range of $t_1$ is within 1 to 20 $fs$.

By considering cold IS, the system, coupled with its environment including the excited (hot NIS) and non-excited molecules, a non-equilibrium initial condition from Eq.(\ref{eq1}) for the rest of its dynamics given by $t >> m/\xi$ can be determined
\begin{eqnarray}
\frac{1}{2} m \langle \vec{v}^2_{s}(0) \rangle = \frac{3}{2} k_B T_>,
\label{eq5}
\end{eqnarray}
where $T_>$ is the local temperature right at the core of the particle track at the non-equilibrium state.
Eq. (\ref{eq5}) is equivalent to the equipartition of the kinetic energy of molecules in an ideal gas.
$T_>$ can be determined by the temperature profile as illustrated in Fig.\ref{fig:Tprofile}.
The subscript $>$ under $T$ implies that the local temperature within the core of the radiation track passing through the medium is greater than the average temperature of the entire medium that is typically at the room temperature, $T\approx 300K$.

In Eq.(\ref{eq5}) we ignore the coupling between IS and water molecules far from the core of the track-structure.
In the next section we incorporate the relaxation of the IS thermal energy and the cooling down processes from $T_>$ to room temperature using a second type of molecular dynamics and coupling through Lennard-Jones potential between IS and water molecules.

Thus an IS created in the track and surrounded by NIS evolves initially under non-equilibrium conditions until reaching a dynamic equilibrium with NIS within a very short time scale $t_1 = m/\xi \approx 1 fs$.
From there, it evolves slowly through the entire medium with an initial kinetic energy proportional to $T_>$.
The transport of IS and the collisions it experiences is a second channel of dissipation that relaxes IS to lower its kinetic energy to a value proportional to the room temperature.
From there, IS diffuses thermally under Brownian motion.
For clearer classification we may rearrange these stages as following:
1) Production of cold IS.
2) Femto-second NIS to IS energy transfer producing hot ($T_>$) IS.
3) Transport/collision of hot IS through the medium until relaxation to room temperature.

The initial velocity, $\vec{v}_s(0)$, is a function of the local temperature, $T_>$, as given by Eq. (\ref{eq5}).
To get an idea of its order of magnitude, let us consider a Hydrogen atom with mass $m_H$
\begin{eqnarray}
v_s(0) &=& \sqrt{\frac{3k_B T_>}{m_H}} = \sqrt{\frac{3k_B T}{m_H}} \sqrt{\frac{T_>}{T}} \nonumber \\
&=& 2,725  \frac{m}{s} \sqrt{\frac{T_>}{300}}.
\label{eq54}
\end{eqnarray}
From Eq.(\ref{eq54}) we infer that increasing $T_>$ up to 10,000K (see Fig.\ref{fig:Tprofile}) results in an increase pf up to a factor 6 in $v_s$.
Note that $v_s$ calculated in Eq.(\ref{eq54}) is greater than the speed of sound in water, which is approximately $1,500 m/s$~[\onlinecite{Sound-wave}].
With $T_> = 10,000 K$, $v_s$ turns out to be approximately equal to $15732 m/s$, which is one order of magnitude higher than the speed of sound.
For that reason, one may expect such large molecular velocities to form nanoscopic shock-waves~[\onlinecite{Zeldovich1996:Book,Sedov1946:PMM,vonNeumann1947:Book,Landau_Lifshitz1987:Hydro_Book}] in the track-structure as discussed in a series of publications (see for example Refs.~[\onlinecite{deVera2019:CN,Surdutovich2010:PRE}]).
%
%

\begin{figure}
\begin{center}
\includegraphics[width=1.0\linewidth]{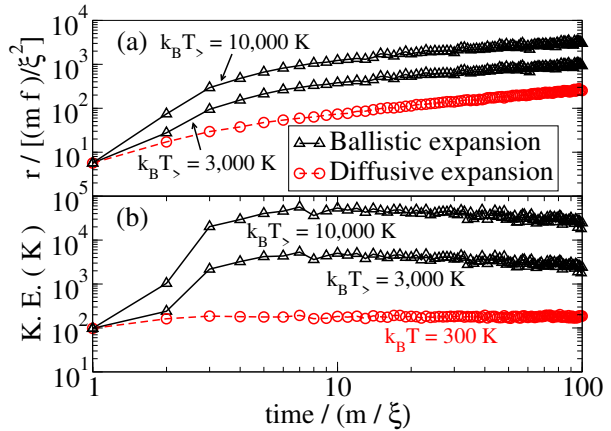}\\ 
\noindent
\caption{
The lateral coordinate, $|\vec{r}|$ (a), and kinetic energy (K.E.) of an IS plotted on a logarithmic scale (b), as a function of time and temperature. At the early stage of the time evolution of IS, where $t \approx m / \xi$, the collective motion is governed by a rapid ballistic expansion that gradually reaches to the thermal equilibrium at room temperature.
For illustration we consider a temperature of NIS, up to $T_> = 10,000$ K.
}
\label{Fig4}
\end{center}\vspace{-0.5cm}
\end{figure}

\section{Spontaneous cooling down processes: an MD simulation}
\label{SecLJ}
Immediately after exit of IS from the hot core of the track structure with an initial velocity given by Eq.(\ref{eq54}), it collides with water molecules which are at the thermal equilibrium in room temperature.
Through a series of collisions, IS loses its kinetic energy in an irreversible process and cools down to room temperature.
IS must travel a distance $\ell$ within the relaxation time $\tau$ to lose its energy and become equilibrated to room temperature.
This subsection is devoted to estimate values of $\ell$ and $\tau$.
For this reason, it is adequate to treat the problem with a Newtonian system of many particles interacting through a Lennard-Jones (LJ) potential.
Because of finite number of particles in MD and to handle the time reversal symmetry of the equations of motion of the microstate, and the Poincare recurrence time~[\onlinecite{Wan2021:PTRS}], we calculate $\tau$ after mixing and averaging ensemble of IS trajectories.

Below we employ a classical MD of LJ particles to calculate the transport coefficients in the condensed phase and calculate the relative expansion of the hot IS compared to cold IS due to a gain in their kinetic energies from NIS of the track structure.
The length corresponding to this expansion can be considered the expansion radius of the burst of tracks, or the shock-wave front at the nanoscopic scale~[\onlinecite{Hafskjold2021:PRE}].
We choose the parameters of the LJ particles (such as mass and their pair-wise interaction parameters) to replicate the typical diffusion constant of radiolysis species, $D = 4.3 \times 10^{-9} m^2/s$, implemented in TOPAS n-Bio.

\subsection{Newtonian system of many particles}
We consider a system of $N$-IS in the track-structure with initial velocities $v_0 = v_s(T_>)$ given by Eq.(\ref{eq54}), interacting with water molecules outside of the track structure through a Newtonian equation of motion
\begin{eqnarray}
m_i\frac{d\vec{v}_i(t)}{dt} = \sum_{j=1}^{N'}\vec{F}_{i,j}(t).
\label{eq1x}
\end{eqnarray}
where $i = 1,2,\dots, N$ runs over the number of IS created by the passage of radiation and $j = 1,2,\dots, N'$ is the total number of water molecules that constitute the thermal bath (water molecules outside of the core of the track); thus $N' >> N$.

The pair-wise LJ potential is defined by
\begin{eqnarray}
V_{LJ}(r) = 4\varepsilon \left[\left(\frac{\sigma}{r}\right)^{12} - \left(\frac{\sigma}{r}\right)^6\right],
\label{eq2x}
\end{eqnarray}
where $r = |\vec{r}_i - \vec{r}_j|$ is the inter-atomic distance, $\varepsilon$ is the depth of the potential well, and $\sigma$ is the distance at which the particle-particle potential energy $V_{LJ}$ is zero.
The LJ potential has its minimum at a distance of $r = r^* = 2^{1/6} \sigma$ where $V_{LJ}(r^*) = -\varepsilon$.
Thus
\begin{eqnarray}
\vec{F}(r) = - \hat{r} \frac{\partial}{\partial r} V_{LJ}(r) =
24\frac{\varepsilon}{r} \left[2\left(\frac{\sigma}{r}\right)^{12} - \left(\frac{\sigma}{r}\right)^6\right]\hat{r},
\label{eq2x}
\end{eqnarray}
is a pair-wise force acting on the LJ particles.

\begin{figure}
\begin{center}
\includegraphics[width=0.8\linewidth]{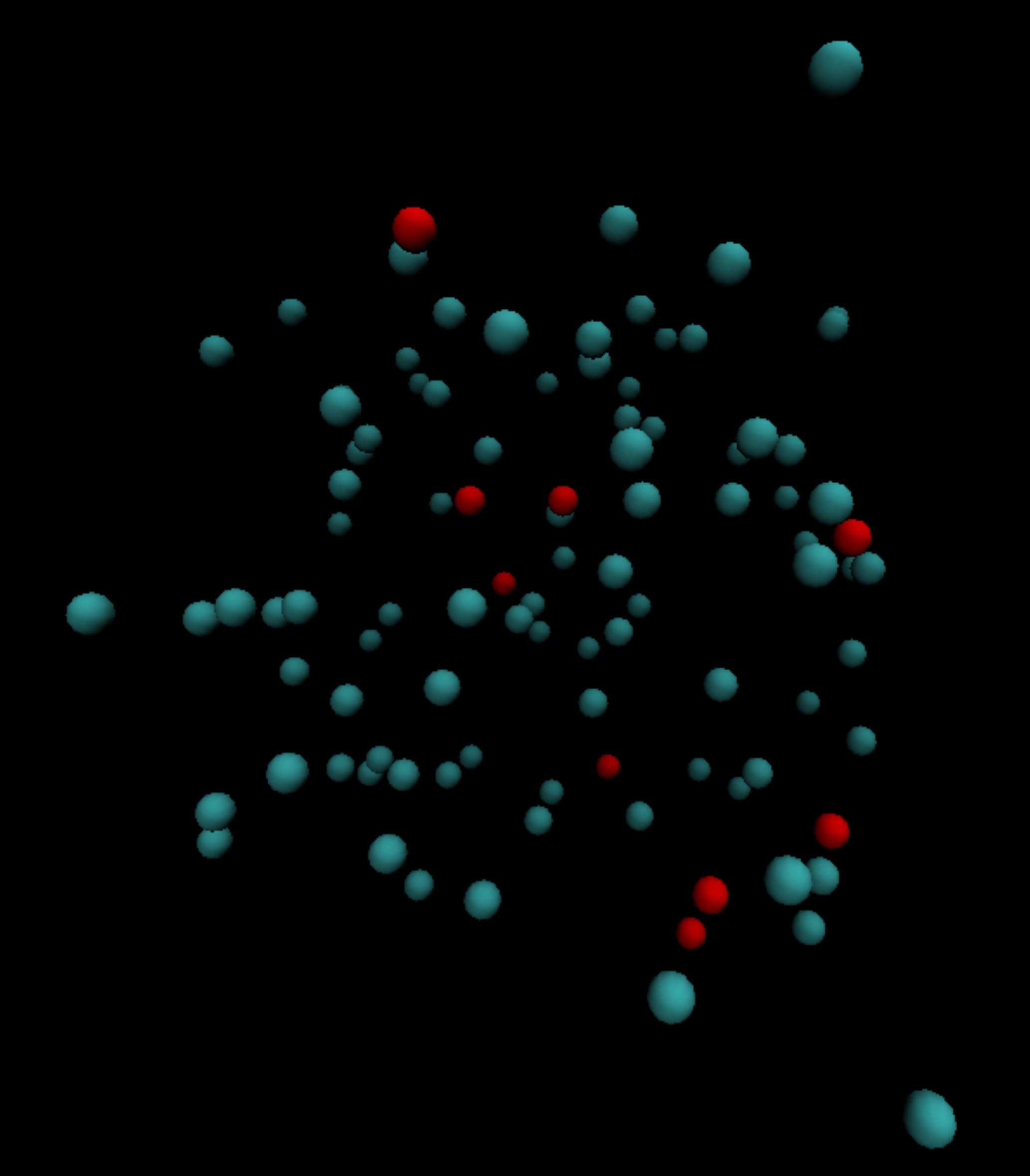}\\ 
\noindent
\caption{
A typical sample of the initial configuration of atoms interacting with the Lennard-Jones (LJ) potential.
The blue spheres are cold atoms in the thermal bath and the red spheres represent hot atoms with initial random velocities one order of magnitude greater than the velocities of the atoms in the thermal bath.
}
\label{Fig4_b}
\end{center}\vspace{-0.5cm}
\end{figure}

\begin{figure}
\begin{center}
\includegraphics[width=1.0\linewidth]{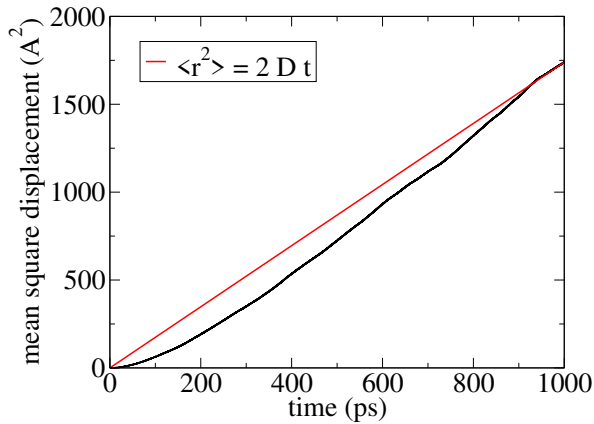}\\ 
\noindent
\caption{
Mean square displacement of room temperature IS as a function of time (black line) obtained from the dynamic trajectories with atoms equilibrated with room temperature.
The red line represents the Einstein relation obtained from interpolation of the MD points (black line). The diffusion constant obtained in this simulation is $D = 8.7\times 10^{-9} m/s$, close to the typical diffusion constant of water radiolysis products, $D_w = 4.3 \times 10^{-9} m/s$.
}
\label{Fig5}
\end{center}\vspace{-0.5cm}
\end{figure}

\begin{figure}
\begin{center}
\includegraphics[width=1.0\linewidth]{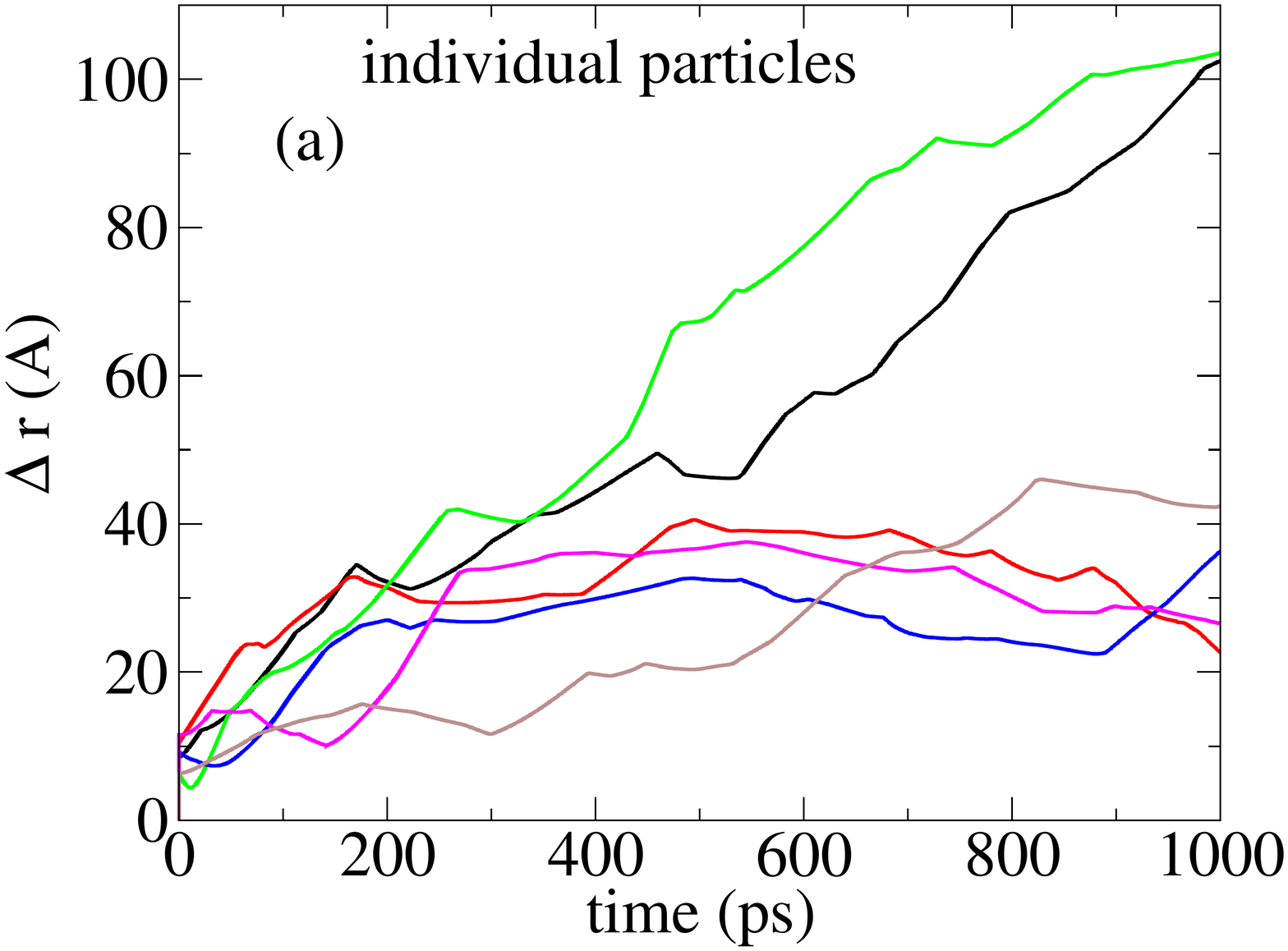}\\ \vspace{-0.3cm}
\includegraphics[width=1.0\linewidth]{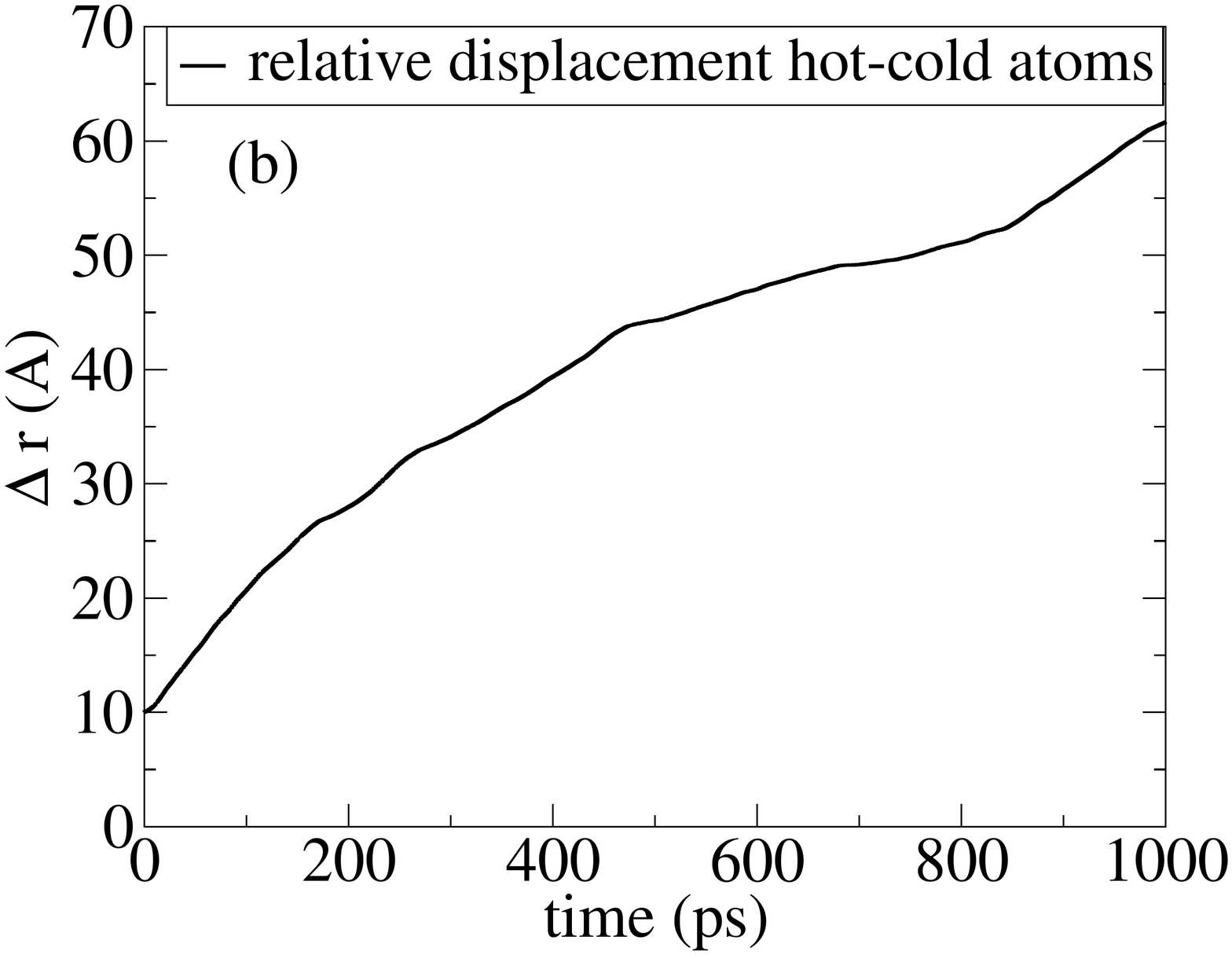}\\ 
\noindent
\caption{
(a) A sample of the instantaneous displacement of four individual hot particles relative to the same particles at room temperature obtained after running two separate MDs with 100 particles interacting through a LJ-potential at a temperature equal to 300 K.
In the first MD, in order to simulate the diffusive motion and calculate the diffusion constant of standard Brownian motion at thermal equilibrium as shown in Fig. \ref{Fig5}, the cold particles are set initially with random positions and velocities.
To simulate the ``hot particles", we picked ten particles from the first MD and scaled their initial velocities by a factor of 10. 
The other 90 particles act as a thermal bath for these 10 hot particles to absorb and dampen their extra kinetic energies.
}
\label{Fig6}
\end{center}\vspace{-0.5cm}
\end{figure}

The physical units in our MD computational codes are set to reduced units such that the Boltzmann constant is scaled to unity, $k_B=1$, where in SI units
%
%
$k_B = 1.381 \times 10^{-27} kg \cdot A^2 \cdot ps^{-2} \cdot K^{-1}$.
Thus we consider Angstrom, ps, and Kelvin as units of length, time and temperature, and express the mass of particle in the reduced mass unit of $m^* = m [kg]/(1.381 \times 10^{-27})$.

Our MD box consists of 100 LJ-atoms in both simulations of diffusive and shock-wave atoms.
In both sets of simulations, $N'=90$ and $N=10$ where $N'$ and $N$ refer to number of NIS and IS.
In the diffusive simulation, the IS have an initial velocity of room temp whereas in the shock-wave their initial velocities scale up to ten times larger than the initial velocities in diffusive simulation.
We applied periodic boundary condition (PBC) with the side length of the cube box that contains LJ particles, equal to 14 Angstrom.

The LJ parameters to replicate the diffusion constant of typical water radiolysis products with mean diffusion constant of $D \approx 4 \times 10^{-9} m/s$, as depicted in Figs. (\ref{Fig5}-\ref{Fig6}), are
$\varepsilon=3000 K$ and $\sigma=2.5 A$.
These are the parameter of LJ potential, the depth of the well and the distance at which the inter-particle potential is 0, respectively.
$r_c=2.5\sigma$ is the cutoff distance that the LJ potential is truncated, i.e., $V_{LJ}(r) = 0$ if $r > r_c$.
Thus the LJ potential at the cutoff distance is
\begin{eqnarray}
V_{LJ}(r_c) = 4\varepsilon \left[\left(\frac{\sigma}{r_c}\right)^{12} - \left(\frac{\sigma}{r_c}\right)^6\right].
\label{eq2x}
\end{eqnarray}
The rest of parameters in our MD are $dt=1 fs$, the MD time step,
$N_{\rm step}=10^6$ is the total running steps in the MD simulation and
$T=300 K$ is the cold-atoms (environment) temperature.

The results of the MD simulation are shown in Fig. \ref{Fig6}(b).
Here $\Delta r$ measures the radial displacement of hot IS wave-fronts compared to cold (room temperature) IS.
Considering the linear growth of $\Delta r$, we may predict
$\Delta r (t) = \Delta r_0 + m_s t$ where $\Delta r_0 = 10 A$ with the slope $m_s = (50 A) / (1000 ps)$.
Thus at $t = 1 \mu s$, we may extrapolate the results to predict $\Delta r (1 \mu s) = 10 A + (50 A) / (1000 ps) \times 10^6 ps \approx 5 \mu m$.
Further simulations using TOPAS n-Bio are on the way to calculate the relaxations to room temperature~[\onlinecite{Baikalov:unpublished}].

\section{Results and Discussions}
\label{SecRes}
To evaluate the diffusive range of IS, and compare it with the corresponding data currently available in several widely-used MC simulations of ionizing radiation track-structures, and examine the significance of the inter-track correlations under FLASH-UHDR conditions, we have conducted a series of computational and analytical studies at nanoscopic scales.
Without loss of generality, the focus of illustrative results is on passage of a scanning beam of protons within a range of energies relevant to their applications in radiotherapy.

The results of our Geant4-DNA MC simulations have revealed the position of NIS that are concentrated and embedded within nano-scale buffers, wrapping around IS.
Thus NIS are the main source of the intensive heat-waves and the burst of IS and their effects on dynamics and the trajectories of IS are not negligible, unlike the current consideration in most practical MC models.

With the assumption of IS-NIS coupling, we have systematically constructed a formalism for a channel of energy transfer such that IS gain a significant amount of kinetic energy from NIS portion of the track.
Our model predicts IS reach distances comparable with the inter-track spacing calculated for a volume receiving 10 Gy at the rate of 40 Gy/s, and higher.

Fig. (\ref{Fig4}) shows the numerical solutions of Eq. (\ref{eq4}) calculated for the motion of IS with an initial condition that is determined by the local temperatures of NIS, given in the legends, on the graphs.
In this figure, the panels (a) and (b) correspond to the lateral displacement and the kinetic energy of an IS, respectively.
$f = \sqrt{2B} = \sqrt{k_B T \xi}$ is the strength of thermal fluctuations with an amplitude that increases by temperature.
Because of this random force, IS gains initial kinetic energy from hot NIS via exchange of phonon fields.
It is evident that with increase in NIS temperature, $T_>$, the radius of ballistic motion initially expands rapidly until the particles reaches a steady state motion outside the core of the track.
From that point IS tends to decay to a diffusive motion and the thermal equilibrium at room temperature because of the collisions with the surrounding water molecules.

This scenario was validated against systems of interacting particles with LJ potential where we have calculated the
relative displacement of the hot-cold particles, denoted by $\Delta r$ in Fig. \ref{Fig6}(b) by MD.
As shown, $\Delta r$ grows linearly as a function of time.
Extrapolation of the MD results to 1 $\mu s$ after the burst of the track, have led to the arrival of the hot-particles to approximately 5 $\mu m$ beyond the position of the same, cold particles.
It appears that 5 $\mu m$ is comparable with the calculated track-spacing under FLASH-UHDR conditions, so the predicted effects in this work has appeared to be significant to the analysis of inter-track interactions at FLASH-UHDR, compared to the same calculation of the particles if they were moving under diffusive motion at room temperature.

More precisely, we have described this mechanism as {\it femto-second spontaneous release of mechanical energy stored initially in NIS within atto-second time scales, to IS, as a result of inverse population of vibrational, rotational, and electronic excitations.
Quantum mechanically, a spontaneous transition through acoustic phonon modes responsible for decay of NIS to kinetic energy of IS takes place with a rate proportional inversely to the rest mass of IS}.
Comparing the transport of IS coupled to NIS, with identical configurations of non-interacting IS at the thermal equilibrium in room temperature, we have demonstrated the energy gain due to hot nano-scale volumes of the track-structures renormalizes the diffusion constant to significantly higher values.

Thus a much higher diffusion constant predicted in the present model may explain the higher chance of inter-track overlaps, hence the higher reaction rates at the FLASH-UHDR.
We therefore provide a speculative justification why research groups relying on the current MC frameworks have reported negligible overlaps, simply because of underestimation of the diffusion constant.
With the inclusion of our corrections, the overlap must be effectively much higher.

\section{Conclusion}
\label{SecConclu}
In this work we have introduced a mechanism for the burst of track-structures immediately after their generations owing to IS-NIS interaction mediated by the quantized field of acoustic phonons and have demonstrated coexistence of ``hot" NIS with ``cold" IS in the radiation track structures.

Our model have started from a series of step-by-step MC simulations such that the spatial coordinates and energy depositions that accounts for the formation of IS and NIS within atto-second time scales was scored using Geant4-DNA toolkits.
In the following step, a sudden expansion and the broadening of the tracks was simulated.
Subsequently we have calculated the inter-track overlaps initiated by non-equilibrium spectrum of high energy IS.
This step were performed using a series of multi-scale MD simulations.
The boundary conditions were obtained from the temperature profiles and geometries of IS and NIS, calculated in our first step of MC track structure simulation.
To this ends, we have derived a stochastic coarse-grained Langevin equation of motion for IS coupled with NIS from the first-principle molecular dynamics.
From this model we were able to describe irreversible femto-second flow of the thermal energy pumping from NIS, mediated by quantized fields of acoustic phonons to IS.

We have compared the predicted range of IS with the mean track-spacings calculated for the deposited dose of 10 Gy and the dose rate of 40 Gy/s, relevant to the FLASH-UHDR experiments, and have reconciled the current results with the recently proposed hypothetical inter-track overlaps with the significant yield of the reactive oxygen species (ROS) aggregation and formation of non-reactive oxygen species (NROS) agglomerates~[\onlinecite{Abolfath2020:MP}].

Conversion of the thermal energy stored in NIS into kinetic energy of IS can be considered as an internal flow of heat in the track-structures that causes the burst.
This mechanism predicts significant corrections to the intra- and inter-track chemical reaction rates at the FLASH-UHDR where the suppression in the former rates compensate with the enhancement of the latter rates.
Thus, the calculated range of the track expansion is much greater than the prediction of the present MC models that simulate the motion of IS by the thermal diffusion at room temperature, ignoring the effects induced by hot NIS.

We therefore recommend, all MC codes developed for the purpose of nanoscopic simulation of the track-structures
such as Geaant4~[\onlinecite{Agostinelli2003:NIMA}], Geant4-DNA~[\onlinecite{Incerti2010:IJMSSC}], TOPAS-nBio~[\onlinecite{Schuemann2019:RR,Faddegon2020:PM}], PARTRAC~[\onlinecite{Friedland2011:MR}], RITRACKS~[\onlinecite{Plante2011:RPD}], and gMicroMC~[\onlinecite{Lai2021:PMB}]
implement the effect of NIS and the intra-track temperature profile in the diffusion of IS to enable them for the analysis of FLASH-UHDR effects.


\end{document}